\documentclass[a4paper,12pt]{article}         
\usepackage{amsmath,amsfonts,amsthm,amssymb}  % Typical maths resource packages
\usepackage[utf8]{inputenc} % Encoding
\usepackage[T1]{fontenc}    % Font encoding
									   
\usepackage{graphicx,caption}            % Packages to allow inclusion of graphics
\usepackage{color}                       % For creating coloured text and background
\usepackage{hyperref}                    % For creating hyperlinks in cross references
\hypersetup{breaklinks=true}

\usepackage{breakurl}
\usepackage{grffile}
\usepackage{makeidx}                  
\usepackage{acronym}
\usepackage{mathrsfs}
\usepackage{float}
\usepackage{vruler}                    
\usepackage{palatino, url, multicol}
\usepackage{setspace}                  % package for changing the line spacing of a document
\usepackage{txfonts}
\usepackage{booktabs}

\usepackage{tikz}
\usepackage{tikz-3dplot}
\usetikzlibrary{shapes.geometric, arrows, calc, 3d}

\usepackage[numbers,sort&compress]{natbib}   % use \citep for compression
\usepackage{import}

\usepackage{caption}
\usepackage{subcaption}
\usepackage{listings}
\usepackage{comment}
\usepackage[title]{appendix}

\newtheorem{theorem}{Theorem}[section]

\newtheorem{corollary}[theorem]{Corollary}

\restylefloat{figure}

\title{Bridging Statistical Scattering and Aberration Theory: Ray Deflection Function - II: Numerical Validation}

\author{Netzer Moriya}
\date{}

\begin{document}

\maketitle

\begin{abstract}
This paper presents a comprehensive experimental validation of a recently developed Ray Deflection Function (RDF) approach, 
which offers a new framework for modeling surface roughness effects in optical systems. Through detailed geometrical ray 
tracing simulations, we demonstrate that the RDF methodology successfully bridges two traditionally separate domains: 
statistical scattering models and deterministic aberration analysis. We implement and compare the two approaches for modeling 
a parabolic mirror with surface imperfections with three cases: (1) an ideal parabolic mirror baseline, (2) the conventional 
Harvey-Shack (HS) statistical scattering theory applied to ray perturbations, and (3) the newly proposed aberration 
term method based on the RDF theory. 
Our results confirm the statistical equivalence between the HS approach and the RDF-based aberration term method, 
with both producing close near-focal-plane distributions and focal volume characteristics. By establishing this equivalence, 
we validate that surface roughness effects can be accurately represented as deterministic aberration terms while maintaining 
fidelity to established statistical scattering models. 
\end{abstract}

\section{Introduction}
Surface roughness significantly impacts the performance of optical systems, causing scattered light that degrades image 
quality and reduces contrast. Traditional approaches to modeling the effects of surface roughness typically follow one 
of two distinct methodologies~\cite{harvey_shack, beckmann_spizzichino, Harvey2007, duparre2002surface}: 
detailed physical modeling of surface topography with direct ray-tracing, or statistical scatter models that operate primarily 
in the angular domain. Both approaches present significant limitations—the former is computationally intensive and requires 
detailed surface data with sophisticated beam-roughness interaction models. The latter, exemplified by Harvey-Shack theory, 
establishes a direct proportional relationship between the bidirectional reflectance distribution function (BRDF) of a rough 
surface and the power spectral density (PSD) of its height variations. While this provides a statistical framework for predicting 
angular light scattering, it remains fundamentally separate from conventional optical design workflows that use aberration 
formulations to handle smooth surface deviations but have no established mechanism for incorporating roughness effects.

In our theoretical paper \cite{moriya2025a}, we introduced a new conceptual framework that recasts surface roughness effects as 
a ``ray deflection function'' (RDF), which can be statistically represented through a modified Zernike-Fourier hybrid approach. 
This transformation allows surface roughness to be seamlessly integrated with other optical aberration terms by expressing its effects 
through equivalent modifications to the ideal mirror shape.

A theoretical analysis justifying the use of ray-based modeling for realistic optical systems with surface roughness is 
provided in \ref{App:GeometricDiffractionEffects}.

The present paper focuses on the numerical validation of this theoretical framework through detailed simulations. We implement 
the RDF approach using statistical perturbations to rays reflected from an ideal parabolic mirror and compare the results with 
conventional Harvey-Shack scatter theory. By demonstrating the statistical equivalence between these approaches, 
we establish that our method maintains the physical accuracy of established scattering models while offering significant 
computational advantages and enhanced integration with traditional optical design frameworks.

\section{Theoretical Foundation}
Prior to presenting our experimental implementation and results, we summarize the key aspects of the RDF theoretical 
framework to provide context for the validation study (see in \cite{moriya2025a} and references therein).

\subsection{The Ray Deflection Function Concept}
We define the Ray Deflection Function (RDF), denoted by $\mathbf{D}(\mathbf{r}_0)$, as the local angular deflection experienced by 
a ray upon reflection due to a height perturbation at the surface point. 
For small perturbations under the scalar wave approximation, this deflection can be expressed in terms of the local phase 
gradient~\cite{moriya2025a}:

\begin{equation}
\mathbf{D}(\mathbf{r}_0) = -\frac{\lambda}{4\pi}\nabla\Phi(x, y)
\end{equation}

where $\Phi(x, y)$ is the phase delay induced by surface height variations, and $\lambda$ is the wavelength of light. This relationship is derived from the eikonal approximation and is valid when surface perturbations are smooth and differentiable, incident and reflected angles are small, and the wavefront phase is well-described by scalar field theory.

The applicability of this geometric ray approach versus diffraction-based modeling for different optical system parameters is 
analyzed quantitatively in \ref{App:GeometricDiffractionEffects}, where we demonstrate that for practical telescope mirror 
specifications, geometric effects dominate diffraction effects by several orders of magnitude.

\subsection{Relationship to Harvey-Shack Theory}
The Harvey-Shack theory \cite{Harvey1995} establishes that the bidirectional reflectance distribution function (BRDF) is 
directly proportional to the power spectral density (PSD) of surface height variations:

\begin{equation}
\text{BRDF}(\alpha_s - \alpha_i, \beta_s - \beta_i) \propto \text{PSD}\left(\frac{\alpha_s - \alpha_i}{\lambda}, \frac{\beta_s - \beta_i}{\lambda}\right)
\end{equation}

where $\alpha$ and $\beta$ are direction cosines, and subscripts $i$ and $s$ refer to incident and scattered directions, respectively.

Our approach leverages this relationship to derive a ray deflection function that produces the same statistical distribution of 
scattered rays as would be produced by the physical surface roughness. The key insight is that for many practical applications, 
the exact physical roughness pattern is less important than its statistical effect on the reflected light distribution.

\subsection{Aberration Term Formulation}
A central contribution of our theoretical framework is the transformation of statistical surface roughness into equivalent 
deterministic aberration terms that directly modify the ideal mirror shape. The magnitude of this effective shape change 
scales with the RMS roughness while preserving the statistical characteristics of the original PSD model.

For a surface with a given PSD, the phase function $\Phi(x, y)$ can be represented using a modified Zernike-Fourier hybrid approach:

\begin{equation}
\Phi(x, y) = \sum_{j=1}^{N} C_j Z_j(x, y)
\end{equation}

where $Z_j(x, y)$ are Zernike polynomials and the coefficients $C_j$ are statistically determined by the spectral overlap 
between the PSD and the Fourier transform of each Zernike term:

\begin{equation}
C_j = \sqrt{\omega_j} \cdot \xi_j
\end{equation}

with $\xi_j$ representing uncorrelated random variables drawn from a standard normal distribution, 
and $\omega_j$ being the spectral weight of each Zernike mode calculated through:

\begin{equation}
\omega_j = \iint \text{PSD}(f_x,f_y) |F_j(f_r,\phi_f)|^2 df_x df_y
\end{equation}

where $F_j(f_r,\phi_f)$ is the Fourier transform of the $j$-th Zernike polynomial.

This approach ensures that the statistical properties of the original surface PSD are preserved in the phase function 
representation. For computational implementation, the series is truncated at $N$ terms based on a variance capture 
criterion, where the cumulative spectral weight reaches a prescribed fraction of the total surface variance:

\begin{equation}
\frac{\sum_{j=1}^{N} \omega_j}{\sigma^2} \geq 1 - \epsilon
\end{equation}

with $\epsilon$ being a small tolerance parameter (typically 0.01-0.05). 
The number of required terms depends on the spatial frequency characteristics of the surface roughness, with $N \approx 2\pi R/l_c$ 
providing a practical guideline for many common PSDs, where $R$ is the aperture radius and $l_c$ is the characteristic 
correlation length.

This formulation allows roughness effects to be treated as deterministic aberration terms that can be directly integrated 
with conventional optical design methods, while maintaining statistical fidelity to the underlying roughness model and 
preserving the spatial correlation structure implied by the original PSD.

\section{Ray-Tracing Simulation Principles}

The ray-tracing simulation employed in this study systematically evaluates the optical performance of modified mirror surfaces 
by numerically tracing light rays through defined optical paths. The underlying framework is structured around the precise 
geometrical representation of mirror surfaces, computation of surface normals, and subsequent calculation of reflected rays. 
The mathematical principles governing this simulation are outlined below.

\subsection{Surface Definition}

The mirror surface $S$ is defined as a rotationally symmetric function incorporating a classical parabolic term augmented by 
higher-order spherical aberrations up to the 12th order\footnote{In what follows we simulate a parabolic mirror's surface so 
$c_{2n,0}=0 \forall n$.}. Explicitly, the surface height function $z(\rho)$ in cylindrical 
coordinates $(\rho, \theta, z)$ is represented as:

\begin{equation}
z(\rho) = \frac{\rho^2}{4f} + \sum_{n=2}^{6} c_{2n,0}\rho^{2n} + \xi(\rho, \theta)
\end{equation}

where $f$ is the focal length of the parabolic reference surface, $\rho = \sqrt{x^2 + y^2}$ denotes the radial coordinate 
from the optical axis, $c_{2n,0}$ are spherical aberration coefficients characterizing deviations from the ideal parabolic 
form, and $\xi(\rho, \theta)$ represents the synthetic surface roughness modeled through a hybrid Zernike-Fourier approach, 
characterized statistically by an RMS height $\sigma$.

\subsection{Surface Normal Calculation}

The unit normal vector $\hat{\mathbf{N}}$ at any surface point $\mathbf{P}(x,y,z)$ is critical for accurately determining 
reflected rays. It is computed from the gradient of the surface $z(x,y)$:

\begin{equation}
\hat{\mathbf{N}}(x,y) = \frac{(-\frac{\partial z}{\partial x}, -\frac{\partial z}{\partial y}, 1)}{\sqrt{1 + \left(\frac{\partial z}{\partial x}\right)^2 + \left(\frac{\partial z}{\partial y}\right)^2}}
\end{equation}

Numerical differentiation via central finite differences is employed to approximate the partial derivatives, ensuring 
computational efficiency and sufficient accuracy:

\begin{equation}
\frac{\partial z}{\partial x} \approx \frac{z(x+h,y)-z(x-h,y)}{2h}, \quad \frac{\partial z}{\partial y} \approx \frac{z(x,y+h)-z(x,y-h)}{2h}
\end{equation}

where $h$ is a small step size optimized for numerical stability and precision.

\subsection{Incident and Reflected Rays}

The simulation assumes incident rays are parallel to the optical axis, described by the incident ray 
vector $\hat{\mathbf{I}} = (0, 0, -1)$. Given the surface normal $\hat{\mathbf{N}}$, the reflected ray direction $\hat{\mathbf{R}}$ is 
computed according to the classical law of reflection:

\begin{equation}
\hat{\mathbf{R}} = \hat{\mathbf{I}} - 2(\hat{\mathbf{I}} \cdot \hat{\mathbf{N}})\hat{\mathbf{N}}
\end{equation}

This formulation ensures the physically accurate reflection of each incident ray on the surface.

\subsection{Random Point Sampling}
Hit points for ray tracing are randomly selected within the mirror's clear aperture using a uniform distribution subject 
to a margin exclusion criterion. The selection process ensures points lie within the defined aperture boundary:

\begin{equation}
(x_i, y_i) \in S \quad \text{such that} \quad d((x_i, y_i), \partial S) > d_{\text{min}}
\end{equation}

where $d_{\text{min}}$ is a user-defined minimum margin to avoid edge anomalies, and $\partial S$ represents the boundary of 
the mirror aperture. 
This approach ensures proper sampling across the mirror surface while avoiding potential numerical issues near the edges. 
The implementation uses a distance-based criterion to identify and exclude points that are too close to the aperture boundary, 
providing robust sampling for both circular and more complex aperture geometries.

\subsection{Geometric Characterization}
To validate the accuracy of the mirror geometry prior to ray tracing, geometric properties including vertex location and aperture 
centroid are calculated via robust iterative methods. 

Edge points are accurately localized using binary search algorithms, and Singular Value Decomposition (SVD) is employed for 
determining the best-fit plane through the aperture circumference points:

\begin{equation}
\mathbf{X}_{\text{circ}} = \{\mathbf{x}_1, \mathbf{x}_2, \dots, \mathbf{x}_N\}, \quad \mathbf{x}_i \in \partial S
\end{equation}

The best-fit plane normal $\mathbf{N}_{\text{plane}}$ is found by minimizing residuals $\delta z_i$:

\begin{equation}
\min \sum_{i=1}^{N}(\delta z_i)^2, \quad \text{where}\quad \delta z_i = \mathbf{N}_{\text{plane}} \cdot \mathbf{x}_i
\end{equation}

The standard deviation of residuals provides a quantitative measure of aperture planarity and serves as a quality check 
before proceeding with the simulation. This ensures that any observed scattering effects in the results can be attributed 
to the intentionally introduced surface roughness rather than geometric modeling errors.

\subsection{Data Generation and Analysis}

The results are compiled into structured data matrices capturing incident and reflected ray vectors, hit points, and surface normals, 
enabling comprehensive statistical and comparative analysis of optical performance under varied surface modifications and conditions.

This rigorous mathematical and computational framework facilitates an accurate and detailed understanding of optical performance 
implications arising from surface modifications in advanced optical systems.

\subsection{General Simulation Setup}
\label{subsec:general_setup}

To validate our theoretical framework, we implemented the Ray Deflection Function (RDF) approach using statistical perturbations to 
rays reflected from an ideal parabolic mirror with optical axis coincident with the $z$-axis. Our experimental setup consists of a 
parabolic mirror with focal length $f = 2.8$ m and aperture diameter $A_p = 0.4$ m.

For the simulation of surface roughness, we set the imperfections RMS to be characterized by 
height $\sigma = 1 \times 10^{-7}$ m (100 nm). 
This parameter was chosen to represent realistic optical surfaces while remaining well within the validity range of our small-angle 
approximation. 
Surface roughness of this magnitude is typical for moderate-quality optical components and falls within practical manufacturing 
tolerances observed in many optical systems\footnote{Surface roughness parameters such 
as $\sigma = 100$ nm and $l_c = 1$ mm may represent realistic values for conventionally 
polished optical surfaces. This combination of RMS height and correlation length is typical for moderate-quality 
optical components and falls within practical manufacturing tolerances observed in many optical systems.}.

All simulations were performed using a custom ray-tracing implementation that allows for different surface representation methods 
and ray perturbation strategies. Statistical analyses were conducted on ensembles of 15,000 rays uniformly distributed across the
mirror aperture. The ray interactions and subsequent propagation were analyzed at multiple planes along the optical axis to 
characterize the reflected rays distribution in the vicinity of the expected focal plane.

\section{Pure Parabolic Mirror - Baseline Analysis}
\label{sec:Pure_Parabolic_Mirror}

As a baseline for evaluating the effects of imperfections introduced on the surface of a smooth mirror, we first simulated an 
ideal parabolic mirror without any surface roughness.

\subsection{Mirror Geometry and Ray Reflection}

\begin{figure}[H]
    \centering
    \includegraphics[width=7.4cm]{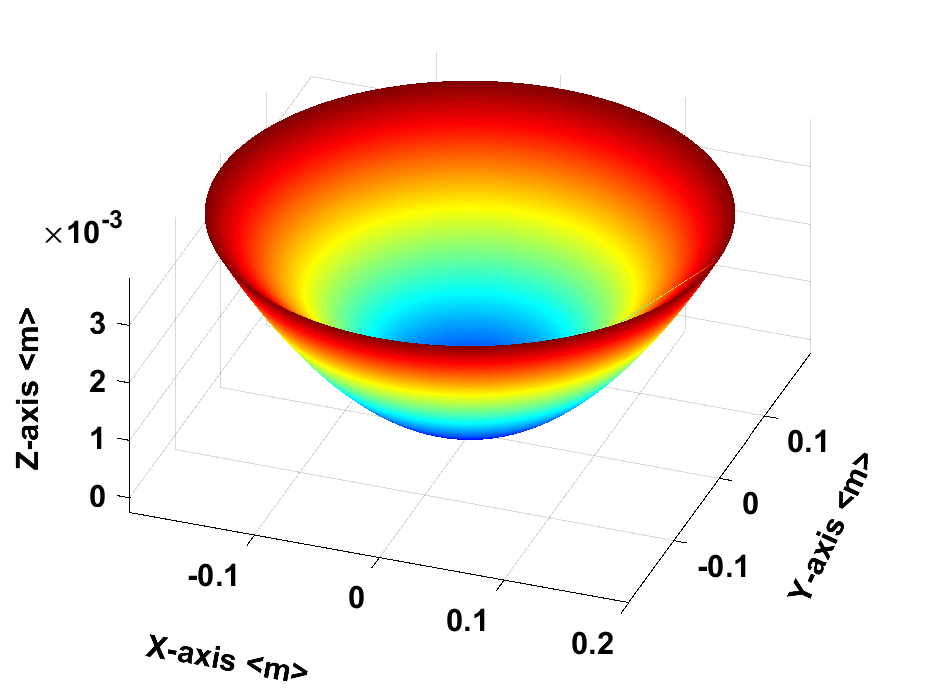}
    \includegraphics[width=7.4cm]{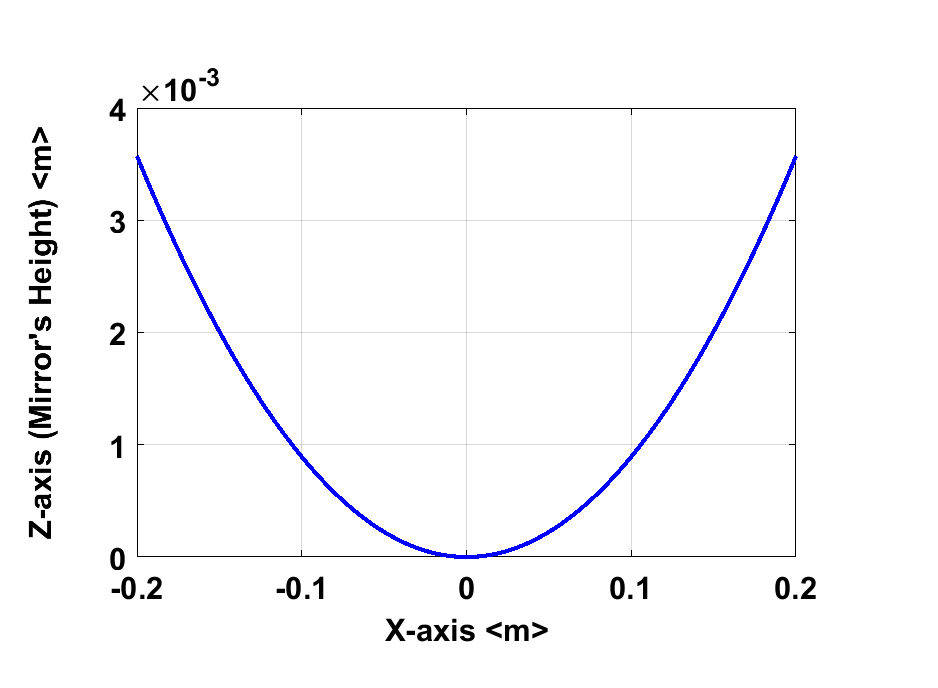}
    \caption{3D visualization of the ideal parabolic mirror surface with focal length $f = 2.8$ m and aperture 
	diameter $A_p = 0.4$ m. The color gradient represents the surface height (left), and 2D profile of the ideal parabolic mirror 
	along the $x$-axis. The profile follows the equation $z(x) = x^2/(4f)$ (right).}
    \label{fig:PAR_mirror_surface}
\end{figure}

The surface of the ideal parabolic mirror follows the equation:

\begin{equation}
z(x, y) = \frac{x^2 + y^2}{4f}
\end{equation}

where $f = 2.8$ m is the focal length. Figure \ref{fig:PAR_mirror_surface} shows the three-dimensional visualization of the 
mirror surface and the two-dimensional profile along the $x$-axis. 
The smooth, continuous surface of the ideal parabolic mirror provides a perfect focus at $z = f$.

\begin{figure}[H]
    \centering
    \includegraphics[width=7.4cm]{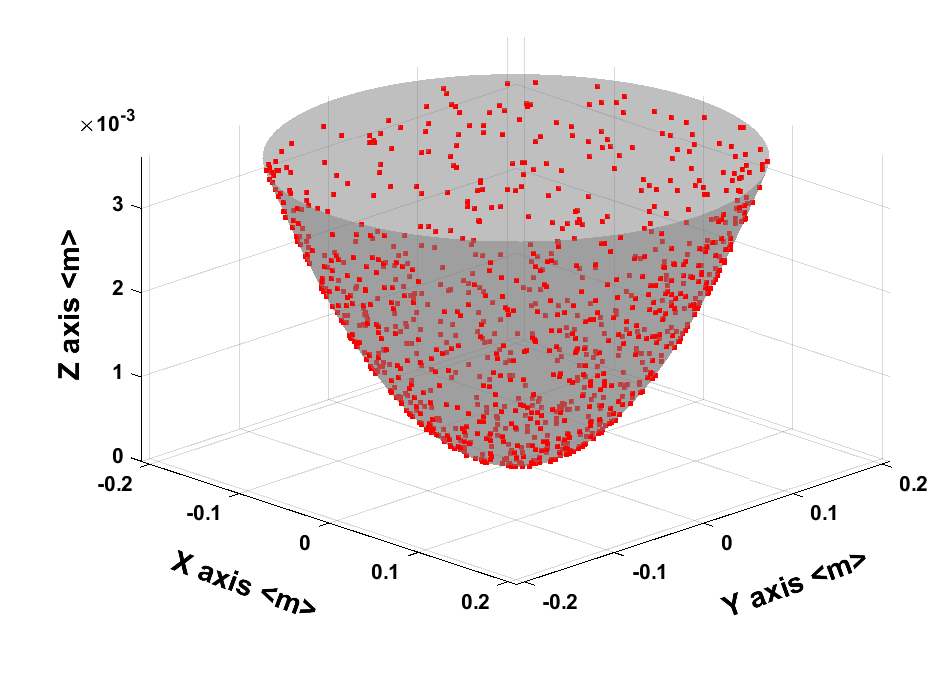}
    \includegraphics[width=7.4cm]{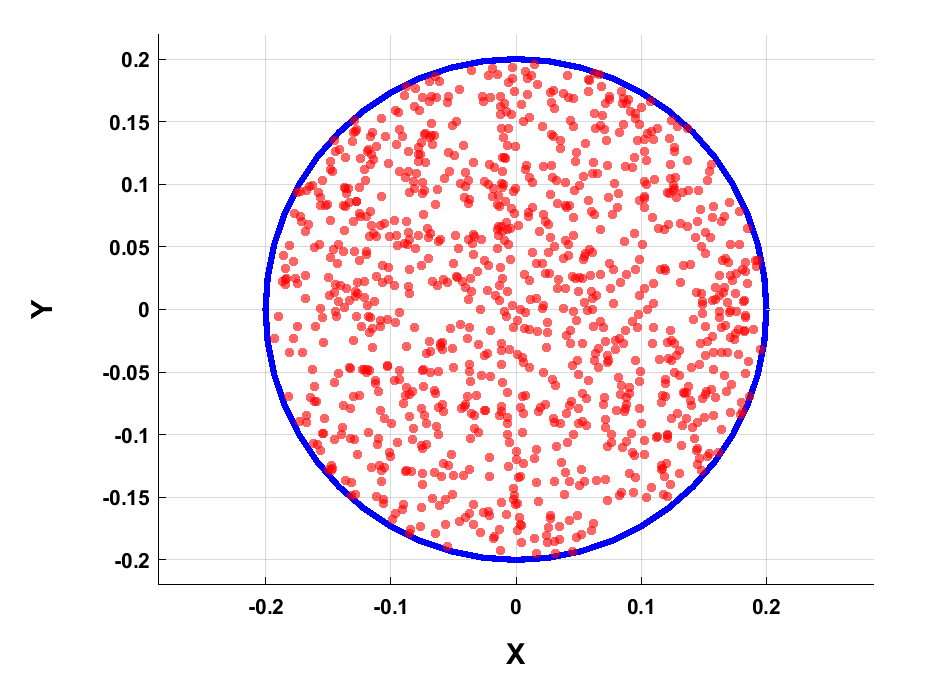}
    \caption{Three-dimensional distribution of randomly selected ray hit points on the mirror surface. 
	The red dots show the uniform distribution of incident ray 
	locations across the mirror's surface (left) and 2D projection of the random ray hit-points on the mirror 
	aperture (right). The uniform distribution ensures comprehensive coverage of the mirror surface for statistical analysis.}
    \label{fig:PAR_Select_Random_3D}
\end{figure}

The points on the mirror's surface where the rays hit and reflected are where selected uniformly in random manner and are
show in figure \ref{fig:PAR_Select_Random_3D}

\begin{figure}[H]
    \centering
    \includegraphics[width=0.9\textwidth]{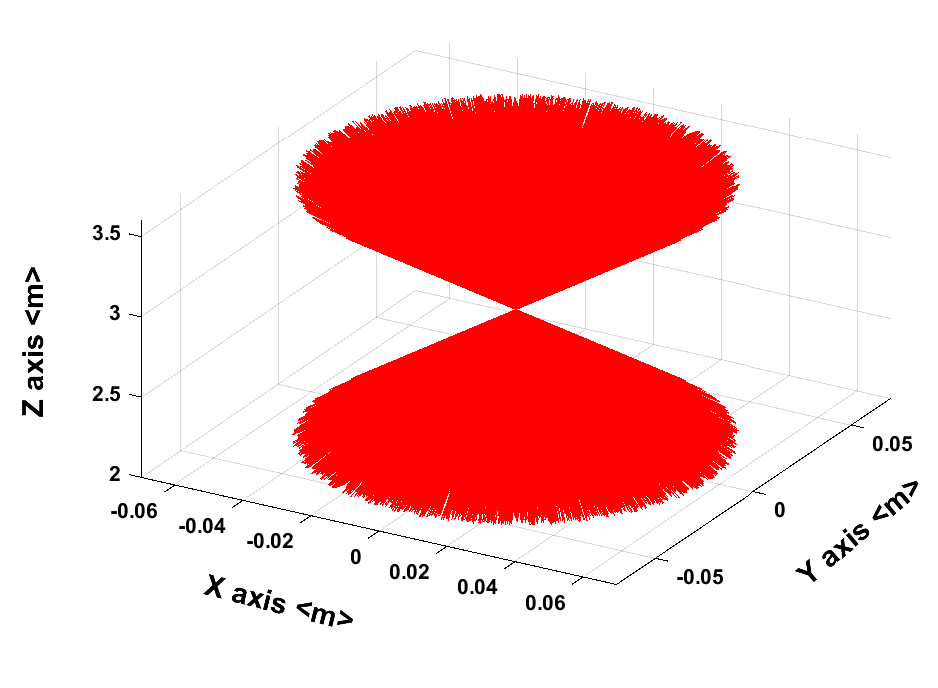}
    \caption{Three-dimensional visualization of the reflected rays from the ideal parabolic mirror. All rays converge at the focal point $z = 2.8$ m, demonstrating perfect focusing behavior with no aberrations.}
    \label{fig:PAR_Reflected_Ray}
\end{figure}

The reflection of incident rays parallel to the optical axis was calculated using the standard law of reflection. 
Figure \ref{fig:PAR_Reflected_Ray} illustrates the reflected ray pattern, with all rays converging precisely at 
the focal point. As expected for an ideal parabolic mirror, the ray convergence is perfect, with no aberrations or 
scattering effects.

\subsection{Focal Volume Analysis}

\begin{figure}[H]
    \centering
    \includegraphics[width=0.9\textwidth]{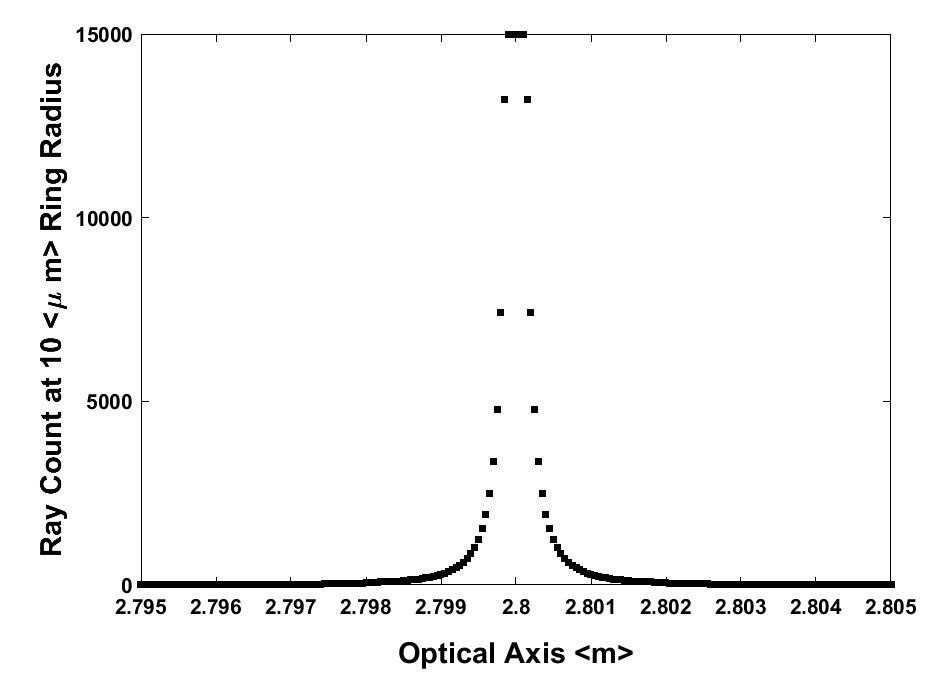}
    \caption{Ray density profile along the optical axis for the ideal parabolic mirror. The plot shows the number of rays 
	contained within a 0.01 mm radius ring at different positions along the $z$-axis. The sharp peak at $z = 2.8$ m indicates 
	perfect focusing with a FWHM of approximately 0.3 mm.}
    \label{fig:PAR_GoThrough}
\end{figure}

To quantify the focusing properties of the ideal mirror, we analyzed the distribution of rays passing through planes perpendicular to the optical axis at various positions near the focal region. Figure \ref{fig:PAR_GoThrough} shows the "ring-through" profile, which represents the number of rays contained within a minimal radius that encloses at least 90\% of the total rays at each $z$ position.

For the ideal parabolic mirror, this profile shows a sharp peak centered precisely at $z = 2.8$ m, with a narrow full width at 
half maximum (FWHM) of approximately 0.3 mm. This represents the geometrically perfect focusing performance of an ideal 
parabolic reflector in the absence of wave effects, and serves as the baseline against which the roughened mirror 
simulations will be compared.

The ray distribution on planar CCD-like grids perpendicular to the optical axis at three positions near the focus is shown in 
Figure \ref{fig:PAR_res}. These images demonstrate the tight concentration of rays at the focal point for the ideal mirror, with minimal spreading as the observation plane moves slightly away from the exact focal position.

\section{Harvey-Shack Approach Implementation}
\label{sec:HS-implementation}

Here, the general setup (see Section \ref{subsec:general_setup}) was applied to implement the Harvey-Shack theory for modeling 
surface roughness effects.

\subsection{Implementation Architecture}

The implementation constructs a statistical representation of the phase gradient field based on the power spectral 
density (PSD) of surface height variations. As established in~\cite{moriya2025a}, the ray deflection function 
is related to the phase gradient by:

\begin{equation}
\mathbf{D}(\mathbf{r}_0) = -\frac{\lambda}{4\pi}\nabla\Phi(x,y)
\end{equation}

For the numerical validation, we employed a uniform PSD bounded by frequency limits $f_\text{min} = 2.5$ cycles/meter 
and $f_\text{max} = 10^6$ cycles/meter, with an RMS roughness value $\sigma = 100$ nm. The optical configuration consists 
of a mirror with aperture $A_p = 0.4$ m and focal length $f = 2.8$ m, operating at wavelength $\lambda = 632.8$ nm.

\subsection{Gradient Field Properties}

\begin{figure}[H]
\centering
\includegraphics[width=0.9\textwidth]{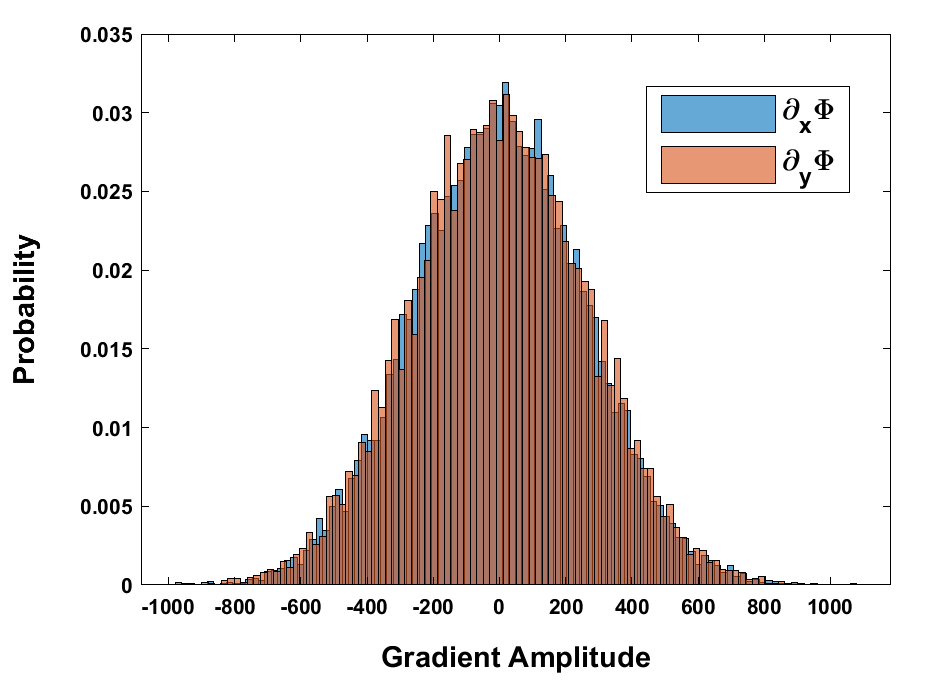}
\caption{Distribution of phase-gradient components $\nabla_x\Phi$ and $\nabla_y\Phi$. Both components exhibit Gaussian distributions with similar statistical properties, confirming the isotropic nature of the gradient field.}
\label{fig:phase_gradient}
\end{figure}

The statistical properties of the phase gradient field are fundamental to the accuracy of the ray deflection model. 
Figure~\ref{fig:phase_gradient} presents a histogram of the x- and y-components of the phase gradient, 
demonstrating their Gaussian distribution with near-identical statistical properties. The measured RMS values 
are $\nabla_x\Phi_\text{RMS} = 244.3$ and $\nabla_y\Phi_\text{RMS} = 253.7$, yielding a ratio of 0.96, which closely 
approximates the expected ratio of 1.0 for an isotropic distribution.

\begin{figure}[H]
\centering
\includegraphics[width=0.9\textwidth]{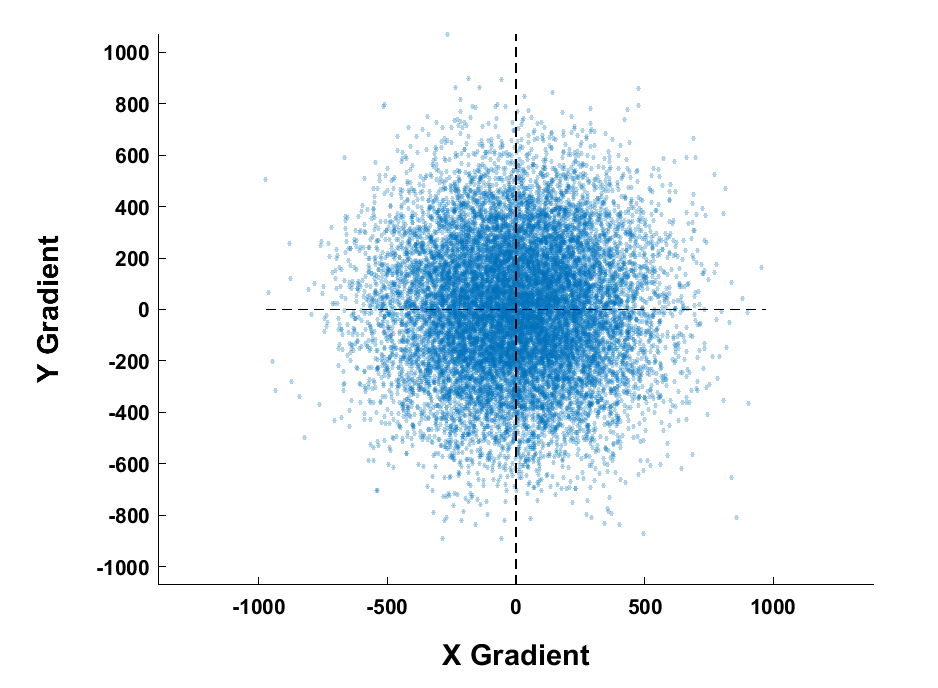}
\caption{Scatter plot of x versus y gradient components with covariance $-1.13 \times 10^3$. The approximately circular 
distribution and low covariance confirm the statistical isotropy of the gradient field.}
\label{fig:xy_gradient_correlation}
\end{figure}

The correlation between gradient components provides additional insight into the isotropy of the simulated field. 
Figure~\ref{fig:xy_gradient_correlation} displays a scatter plot of $\nabla_x\Phi$ versus $\nabla_y\Phi$, revealing an 
approximately circular distribution. The measured covariance between components is $-1.13 \times 10^3$, which represents 
less than 2\% of the variance of either component. This minimal correlation confirms the absence of systematic directional 
bias in the gradient field, as required by the statistical equivalence criteria established in 
Section~\ref{sec:statistical_equivalence}.

\subsection{Phase Function Generation}

The implementation generates a phase function $\Phi(x, y)$ with appropriate spatial frequency characteristics, 
computes its gradient, and applies the ray deflection formula $\mathbf{D}(\mathbf{r}_0) = -\frac{\lambda}{4\pi}\nabla\Phi(x, y)$ 
to obtain perturbed ray directions. The phase function was constructed as:

\begin{equation}
\Phi(x, y) = \sum_{p,q} \sqrt{P_0} \cos(2\pi f_p x + 2\pi f_q y + \phi_{p,q})
\end{equation}

where $(f_p, f_q)$ are spatial frequency components, $\phi_{p,q}$ are random phases uniformly distributed in $[0, 2\pi)$, 
and $P_0$ represents the constant power value per unit area in frequency space, determined by the RMS roughness $\sigma$.

\subsection{Optical Axis Analysis}

\begin{figure}[H]
    \centering
    \includegraphics[width=0.9\textwidth]{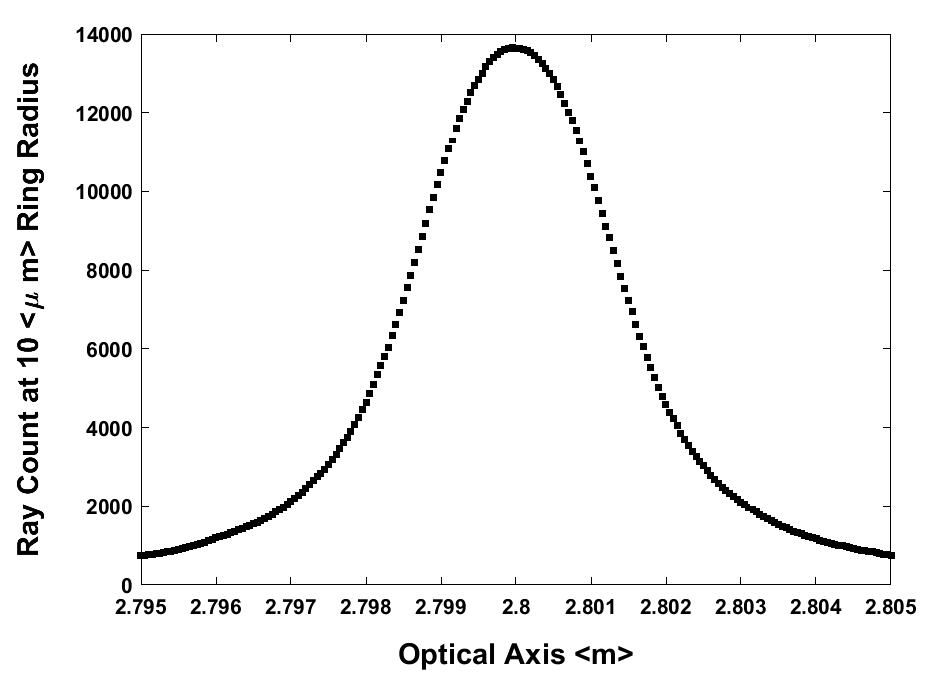}
    \caption{Ray density profile along the optical axis for the mirror with roughness modeled using the Harvey-Shack approach. The plot shows the number of rays contained within a 76 $\mu$m radius ring at different positions along the $z$-axis. The broadened peak indicates the effects of surface roughness on the focal spot, with a FWHM of approximately 2.9 mm, significantly larger than the ideal case.}
    \label{fig:GoThrough_HS}
\end{figure}

As a preliminary overview of the simulation analysis we calculated the density profile of the reflected rays along the optical axis. 
Figure~\ref{fig:GoThrough_HS} presents the number of rays included in a ring with the minimal radius (in this case of 76 $\mu$m) 
that enclosed at least 90\% of the total number of rays involved.

The broadening of the profile is clearly demonstrated around the focal point as compared to the ideal (parabolic) mirror and 
similar to the Aberration-Term approach. The FWHM increased from 0.3 mm for the ideal mirror to 2.9 mm for the HS implementation, 
indicating significant focal spot degradation due to surface roughness effects.

\subsection{Frequency Domain Analysis}
\label{subsec:frequency-sampling}

When sampling a PSD that spans multiple orders of magnitude in frequency space, logarithmic sampling provides efficient coverage. 
For a 2D frequency domain with radial symmetry, the mathematical basis for the 1/r point density relationship can be derived 
as follows:

In linear frequency space $(f_r,\theta)$, the differential area element is:
\begin{equation}
dA_{linear} = f_r df_r d\theta
\end{equation}

For logarithmic sampling, we want equal area elements in logarithmic space. Setting $\xi = \log f_r$, the area element 
in log-radial coordinates becomes:
\begin{equation}
dA_{log} = d\xi d\theta
\end{equation}

Since $d\xi = df_r/f_r$, we have:
\begin{equation}
dA_{log} = \frac{df_r}{f_r} d\theta
\end{equation}

To ensure uniform sampling in log space (constant $dA_{log}$), the point density $\rho(f_r)$ in linear space must satisfy:
\begin{equation}
\rho(f_r) \cdot f_r df_r d\theta = \text{constant}
\end{equation}

Solving for $\rho(f_r)$:
\begin{equation}
\rho(f_r) \propto \frac{1}{f_r}
\end{equation}

This $1/r$ relationship ensures that each logarithmic interval contains approximately the same number of sampling points, providing 
proper coverage across all frequency scales relevant to surface roughness characterization.

\begin{figure}[H]
\centering
\includegraphics[width=7.4cm]{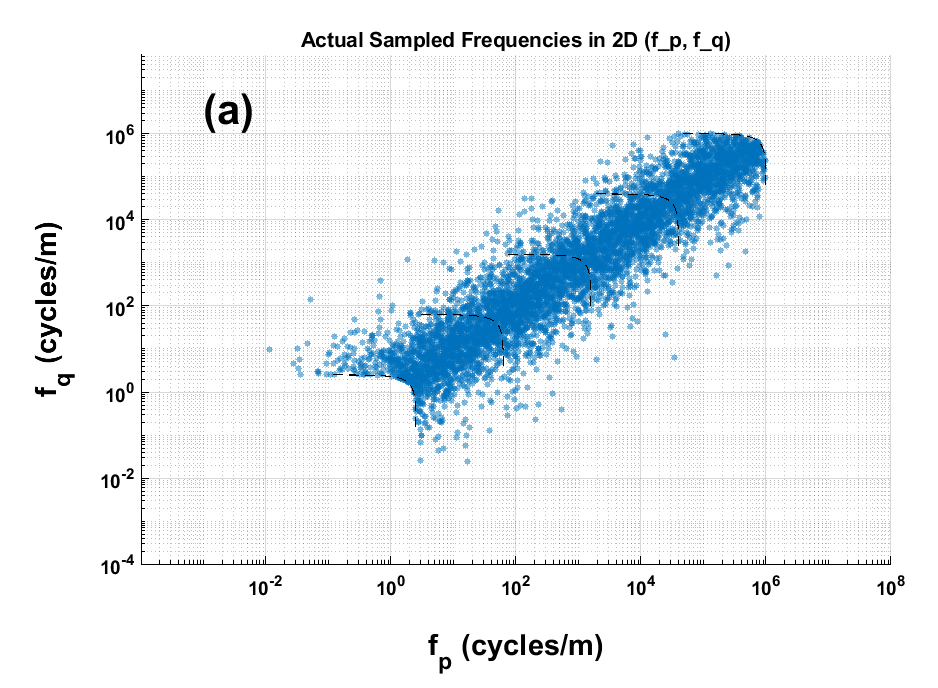}
\includegraphics[width=7.4cm]{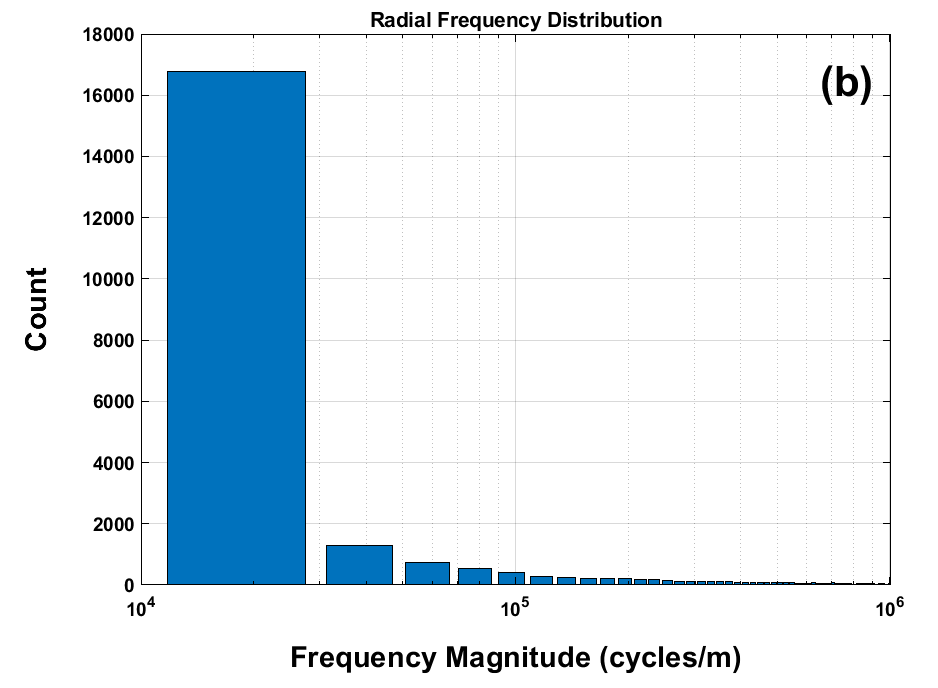} \\
\includegraphics[width=7.4cm]{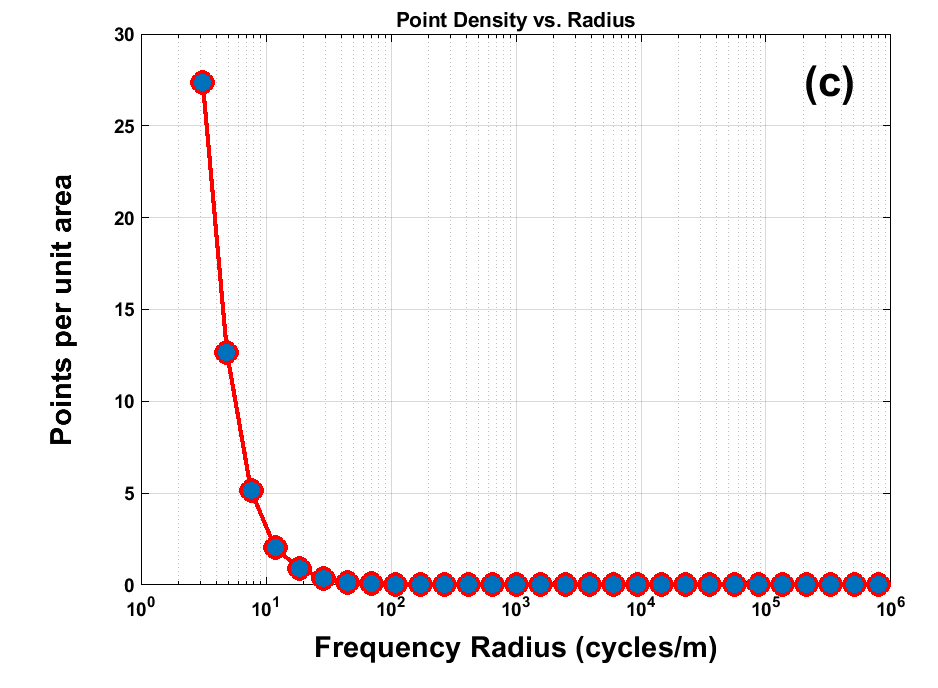}
\includegraphics[width=7.4cm]{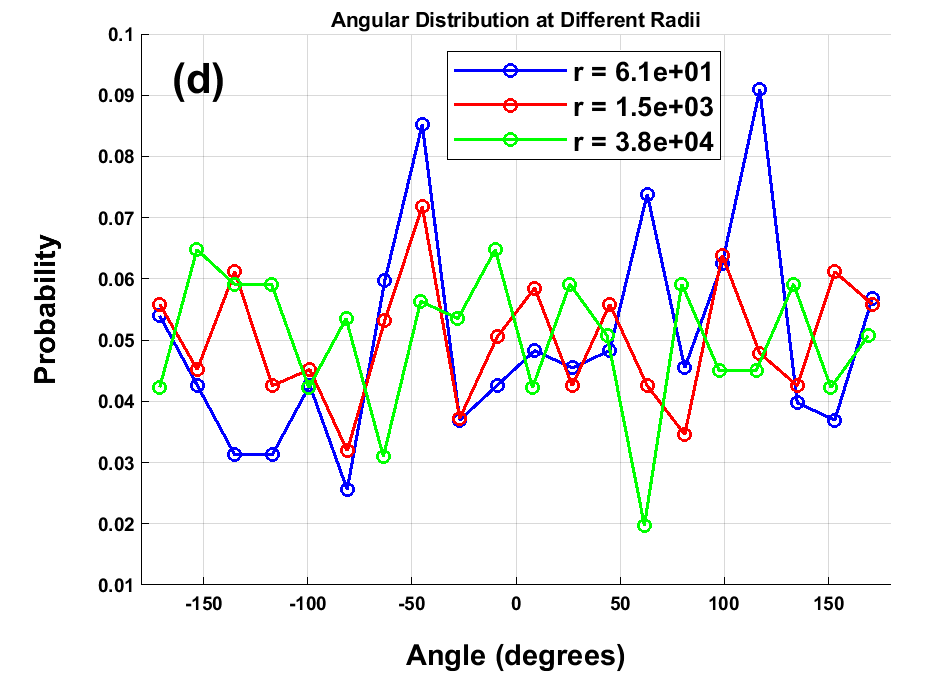}
\caption{Frequency domain sampling characteristics. 
(a) Distribution of sampled frequencies in the $(f_p, f_q)$ plane showing logarithmic coverage from $f_\text{min} = 2.5$ to $f_\text{max} = 10^6$ cycles/meter. 
(b) Radial frequency distribution showing expected concentration at lower frequencies. 
(c) Point density vs. radius exhibiting $1/f$ behavior appropriate for logarithmic sampling. 
(d) Angular distribution at different radii confirming uniform directional sampling.}
\label{fig:frequency_sampling}
\end{figure}

Figure~\ref{fig:frequency_sampling} shows the frequency domain characteristics of our implementation. 
The spatial frequency sampling is performed logarithmically within the 
bounds $f_{min} = 1/A_p = 2.5$ cycles/m and $f_{max} = 10^6$ cycles/m. 

To ensure appropriate sampling across the spatial frequency spectrum, we employed a logarithmic sampling strategy. For a 
uniform PSD within the band-limited region defined by:

\begin{equation}
\text{PSD}(f_x, f_y) = 
\begin{cases}
P_0, & \text{if } f_{\text{min}} \leq \sqrt{f_x^2 + f_y^2} \leq f_{\text{max}} \\
0, & \text{otherwise}
\end{cases}
\end{equation}

The value of $P_0$ is determined by the surface variance $\sigma^2$:

\begin{equation}
P_0 = \frac{\sigma^2}{\pi(f_{\text{max}}^2 - f_{\text{min}}^2)}
\end{equation}

Figure~\ref{fig:frequency_sampling} illustrates the distribution of sampled frequencies in the $(f_p, f_q)$ plane, 
demonstrating effective coverage across the prescribed frequency band. The point density versus radius plot confirms 
the expected $1/f$ behavior for uniform sampling in logarithmic space, ensuring that both low and high spatial frequencies 
are appropriately represented in the model.

The 2D distribution (Fig.~\ref{fig:frequency_sampling}a) confirms uniform angular coverage across logarithmically spaced radii. 
The radial frequency distribution (Fig.~\ref{fig:frequency_sampling}b) shows appropriate concentration at lower frequencies, 
consistent with logarithmic sampling. The point density plot (Fig.~\ref{fig:frequency_sampling}c) demonstrates the 
expected $1/r$ relationship, ensuring proper coverage of the frequency domain. Angular distribution 
analysis (Fig.~\ref{fig:frequency_sampling}d) confirms azimuthal isotropy across multiple frequency bands.

It is important to note that while the sampling density exhibits favoritism toward lower frequencies, this does not contradict the uniform PSD requirements of Harvey-Shack theory. The apparent bias is a computational optimization rather than a statistical one. In Harvey-Shack theory with a uniform PSD, each frequency component within the band ($f_{min}$ to $f_{max}$) should contribute equally per unit area in frequency space, with the PSD defined as:

\begin{equation}
\textrm{PSD}(f_r) = 
\begin{cases} 
P_0, & \textrm{if}\ f_{min} \leq f_r \leq f_{max} \\
0, & \textrm{otherwise}
\end{cases}
\end{equation}
\label{eq:P_0}

where $P_0$ represents the constant power value per unit area in frequency space. For a surface with RMS 
roughness $\sigma$, the value of $P_0$ can be derived from the condition that the total integrated power equals $\sigma^2$:

\begin{equation}
P_0 = \frac{\sigma^2}{f_{max} - f_{min}}
\end{equation}

As detailed above, our implementation uses logarithmic sampling to efficiently cover a frequency range spanning multiple orders 
of magnitude while preserving this statistical uniformity. 
This is achieved through careful amplitude scaling as follows: 
for each sampled frequency point $(f_p, f_q)$ with radial component $f_r$, we calculate the corresponding differential area 
element $df_{element} = f_r \cdot df_r \cdot d\theta$, where $df_r \approx f_r \cdot (log_{10}f_{max} - log_{10}f_{min})/\sqrt{N}$ 
and $d\theta \approx 2\pi/\sqrt{N}$ for $N$ points. 
The amplitude of each component is then scaled proportionally 
to $\sqrt{P_0 \cdot df_{element}}$, ensuring that despite the non-uniform sampling density, the statistical power contribution 
remains uniform across all frequencies. 
In our implementation with $\sigma = 1 \times 10^{-7}$ m, $f_{min} = 2.5$ cycles/m, 
and $f_{max} = 1 \times 10^6$ cycles/m, this yields $P_0 = 1 \times 10^{-20}$ m$^2$/(cycles/m)$^2$. This approach maintains 
fidelity to Harvey-Shack theory while optimizing computational efficiency.

\subsection{Deflection Magnitude and Angular Distribution}

\begin{figure}[H]
\centering
\includegraphics[width=0.9\textwidth]{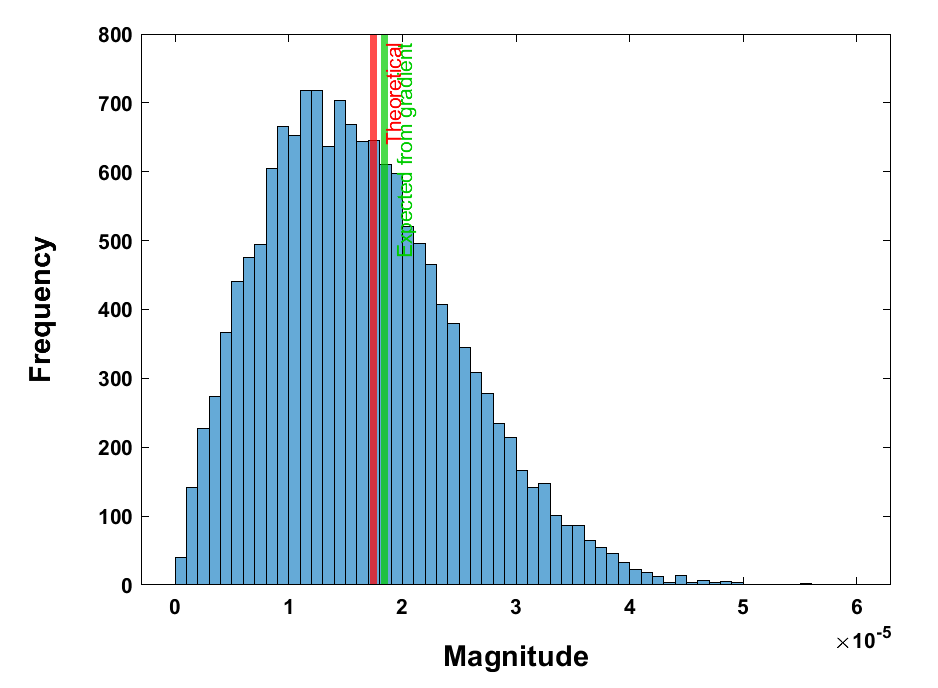}
\caption{Distribution of ray deflection vector magnitudes. The observed mean ($1.57 \times 10^{-5}$) closely matches the theoretical prediction ($1.70 \times 10^{-5}$, dashed line) based on gradient RMS values.}
\label{fig:deflection_magnitude}
\end{figure}

The statistical distribution of ray deflection magnitudes provides a critical validation metric for the RDF approach. 
Figure~\ref{fig:deflection_magnitude} presents a histogram of the computed deflection vector magnitudes. 
The observed mean magnitude ($1.57 \times 10^{-5}$) demonstrates excellent agreement with the theoretical 
prediction ($1.70 \times 10^{-5}$), with a scaling factor of 0.926. This close correspondence validates the 
implementation of the ray deflection formula and confirms proper projection of the gradient onto the tangent plane of reflection.

\begin{figure}[H]
\centering
\includegraphics[width=0.9\textwidth]{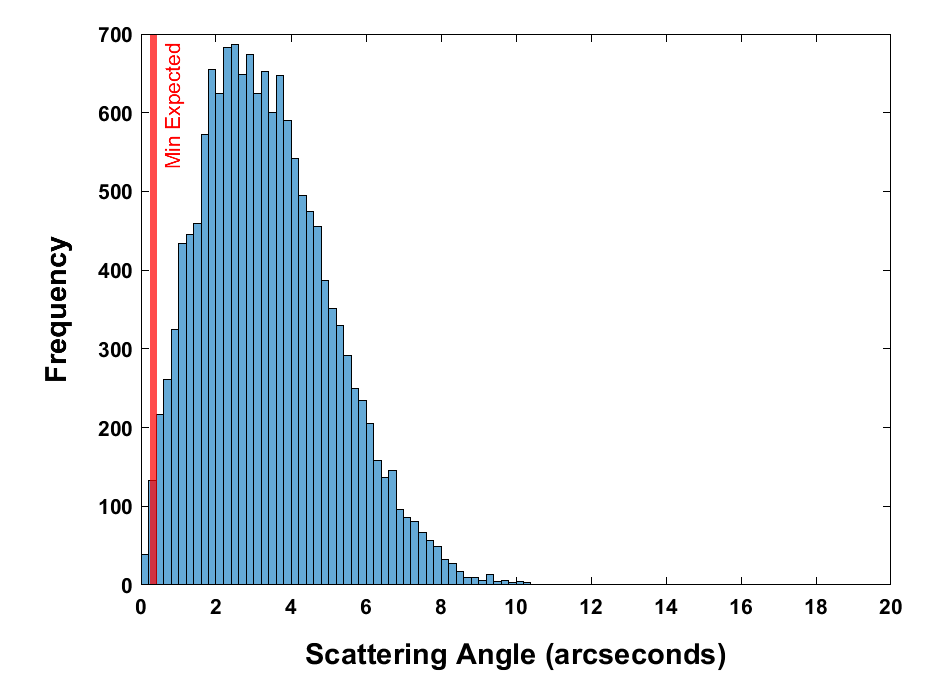}
\caption{Detailed view of scattering angle distribution in the 0-20 arcsecond range. The concentration of scatter angles in 
this region reflects the statistical dominance of lower spatial frequencies in the PSD model.}
\label{fig:scatter_angles_detail}
\end{figure}

The angular distribution of scattered rays, presented in Figure~\ref{fig:scatter_angles_detail}, 
demonstrates the expected behavior. According to Harvey-Shack theory, the BRDF is directly proportional to the PSD:

\begin{equation}
\mathrm{BRDF}(\alpha_s-\alpha_i,\;\beta_s-\beta_i)
\;\propto\;
\mathrm{PSD}\!\Bigl(\tfrac{\alpha_s-\alpha_i}{\lambda},\;\tfrac{\beta_s-\beta_i}{\lambda}\Bigr)
\end{equation}

For a uniform PSD, this corresponds to a maximum scatter angle of 
approximately $\theta_\text{max} = \sin^{-1}(\lambda f_\text{max}) \approx 39^{\circ}$ (141,325 arcseconds). 
While the theoretical range is quite large, the observed distribution is concentrated primarily within 0-8 arcseconds, 
consistent with the statistical dominance of lower spatial frequencies in the deflection process.

\subsection{Statistical Bias Analysis}

\begin{figure}[H]
\centering
\includegraphics[width=0.8\textwidth]{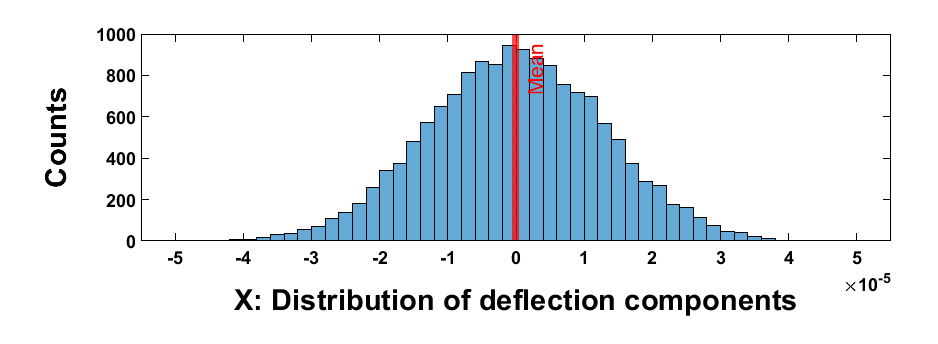} \\
\includegraphics[width=0.8\textwidth]{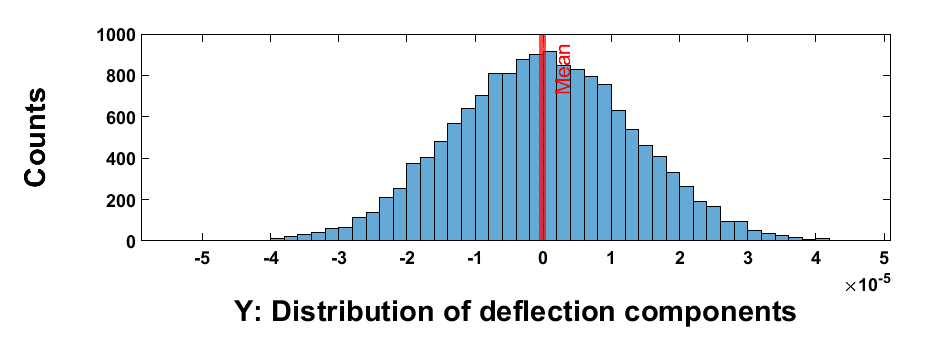} \\
\includegraphics[width=0.8\textwidth]{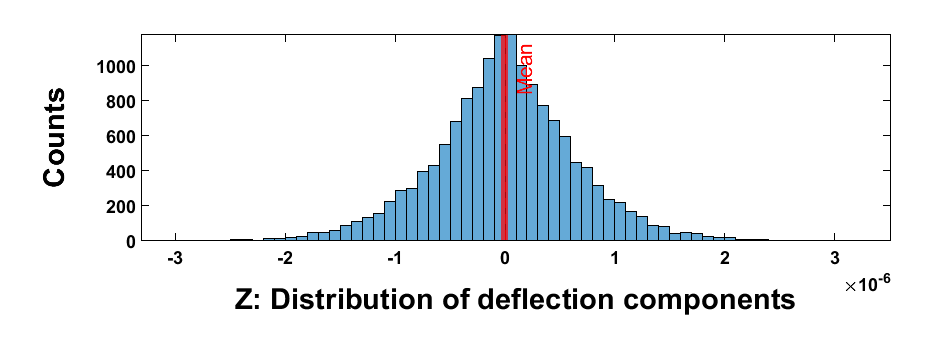} 
\caption{Distribution of x, y, and z components of the deflection vectors. All components exhibit zero-centered Gaussian 
distributions with standard deviations of $1.23 \times 10^{-5}$, $1.28 \times 10^{-5}$, and $6.37 \times 10^{-7}$ respectively, appropriately scaled according to their geometric contributions to the overall deflection.}
\label{fig:deflection_components}
\end{figure}

To ensure the absence of systematic directional bias in the ray perturbations, we examined the mean deflection vector and its 
components. For a statistically valid implementation, the following criterion must be satisfied:

\begin{equation}
P(\Delta\alpha, \Delta\beta) \propto \text{BRDF}(\Delta\alpha, \Delta\beta)
\end{equation}

where $P(\Delta\alpha, \Delta\beta)$ is the probability distribution of ray perturbations. The computed mean deflection 
vector $[-7.41 \times 10^{-8}, 1.47 \times 10^{-7}, -1.48 \times 10^{-10}]$ and the corresponding bias ratio (0.01045) 
indicate negligible systematic bias. Figure~\ref{fig:deflection_components} presents the distributions of the x, y, 
and z components of the deflection vectors, demonstrating balanced, zero-centered distributions in all dimensions.

\subsection{Physical Consistency Analysis}

The RDF implementation produces physically consistent results when compared with established theoretical relationships. The 
measured mean scattering angle of 3.24 arcseconds aligns with predictions based on the relationship between RMS roughness, 
correlation length, and wavelength. For a uniform PSD with bandwidth $[f_\text{min}, f_\text{max}]$, the RMS slope is approximately:

\begin{equation}
\sigma_\text{slope} \approx \sigma \sqrt{\frac{f_\text{max}^2 + f_\text{min}f_\text{max} + f_\text{min}^2}{3}}
\end{equation}

The resulting scattering angle is approximately $\theta_\text{scatter} \approx 2\sigma_\text{slope}$. Notably, the observed 
deflection magnitudes differ significantly from the simple physical approximation $2\sigma/l_c = 2 \times 10^{-4}$, indicating 
that the full PSD-based approach provides a more nuanced prediction than simplified models based solely on RMS roughness and 
correlation length.

\begin{figure}[H]
\centering
\includegraphics[width=0.9\textwidth]{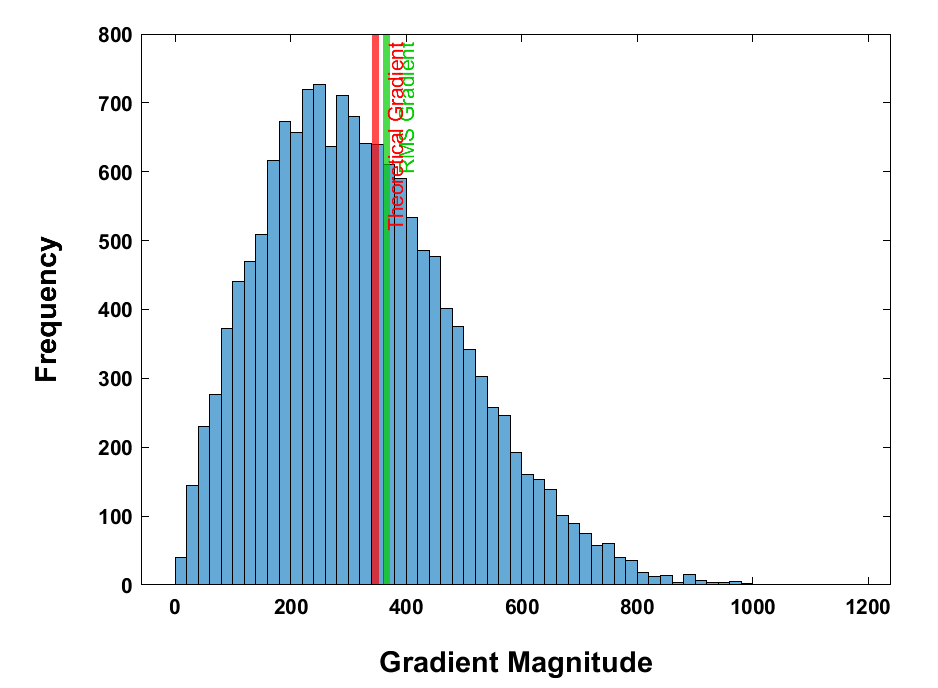}
\caption{Distribution of gradient magnitudes with theoretical RMS value indicated by the dashed line. 
The measured gradient RMS of 352.2 closely matches the theoretical prediction of 337.4.}
\label{fig:gradient_magnitude}
\end{figure}

The coherent relationship between gradient magnitude (Figure~\ref{fig:gradient_magnitude}) and deflection 
magnitude (Figure~\ref{fig:deflection_magnitude}) further validates the physical accuracy of the implementation, 
with the proportionality constant $\lambda/(4\pi)$ correctly scaling between domains.

The numerical implementation demonstrates that the Ray Deflection Function approach successfully bridges statistical scattering 
theory and deterministic ray tracing. The simulated deflection statistics show high fidelity to theoretical predictions while 
maintaining computational efficiency. By properly projecting the gradient onto the tangent plane of reflection, the 
implementation achieves a statistically isotropic scattering pattern that correctly preserves the relationship between 
surface PSD and angular scatter distribution.

The observed statistical properties confirm that the implementation correctly addresses potential artifacts (such as directional bias) 
that can arise in numerical implementations. The close agreement between observed and theoretical deflection statistics validates the 
mathematical framework presented in earlier sections and demonstrates its practical applicability for optical system modeling.

\section{The Aberration Term Implementation}

Instead of perturbing the rays after reflection from an ideal surface (as in the HS approach), the aberration term method directly modifies the mirror profile to incorporate the statistical effects of roughness. The modified surface equation becomes:

\begin{equation}
z_{aberrated}(x, y) = \frac{x^2 + y^2}{4f} + h_{imp}(x, y)
\end{equation}

where $h_{imp}(x, y)$ represents the height perturbation due to roughness. This perturbation field was synthesized to have the same statistical properties as the phase function used in the HS implementation, but translated into physical height variations.

The aberrated mirror profile is shown in Figure \ref{fig:AT_Imperfection_Component_profile}, displaying the imperfection component added to the ideal parabolic shape:

\begin{figure}[H]
    \centering
    \includegraphics[width=0.9\textwidth]{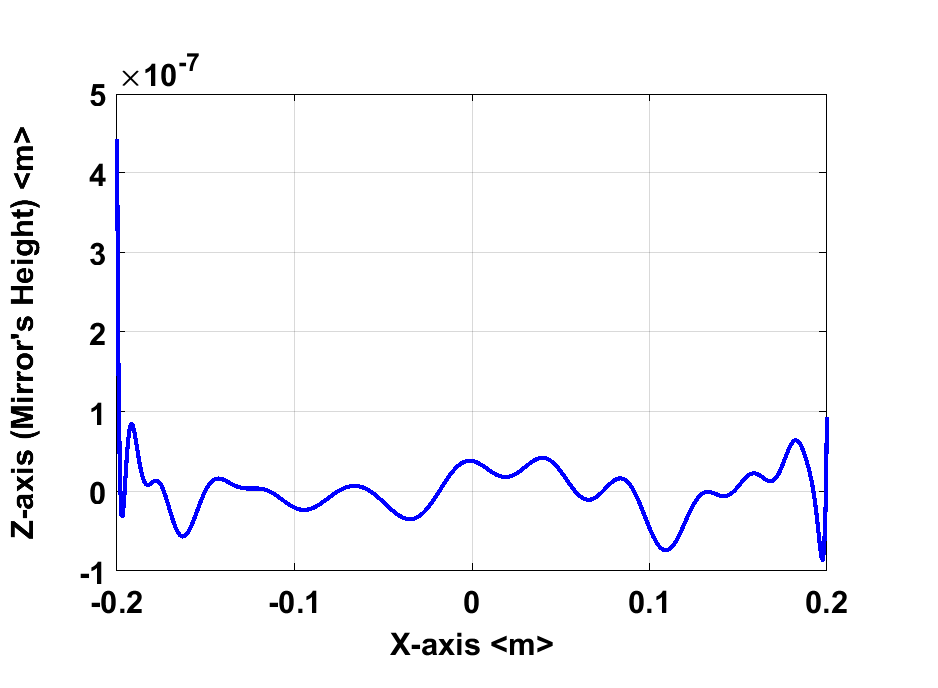}
    \caption{Profile of the imperfection component added to the ideal parabolic mirror shape. The RMS height of this component is 100 nm, corresponding to the specified surface roughness parameter. The profile shows the characteristic statistical variations expected from the uniform PSD model.}
    \label{fig:AT_Imperfection_Component_profile}
\end{figure}

Rays were traced through the system using the modified mirror profile, with reflections calculated according to the local surface normal at each hit point. This approach inherently accounts for the roughness effects without requiring post-reflection perturbations.

The "ring-through" profile for the aberration term implementation is shown in Figure \ref{fig:AT_GoThrough}:

\begin{figure}[H]
    \centering
    \includegraphics[width=12cm]{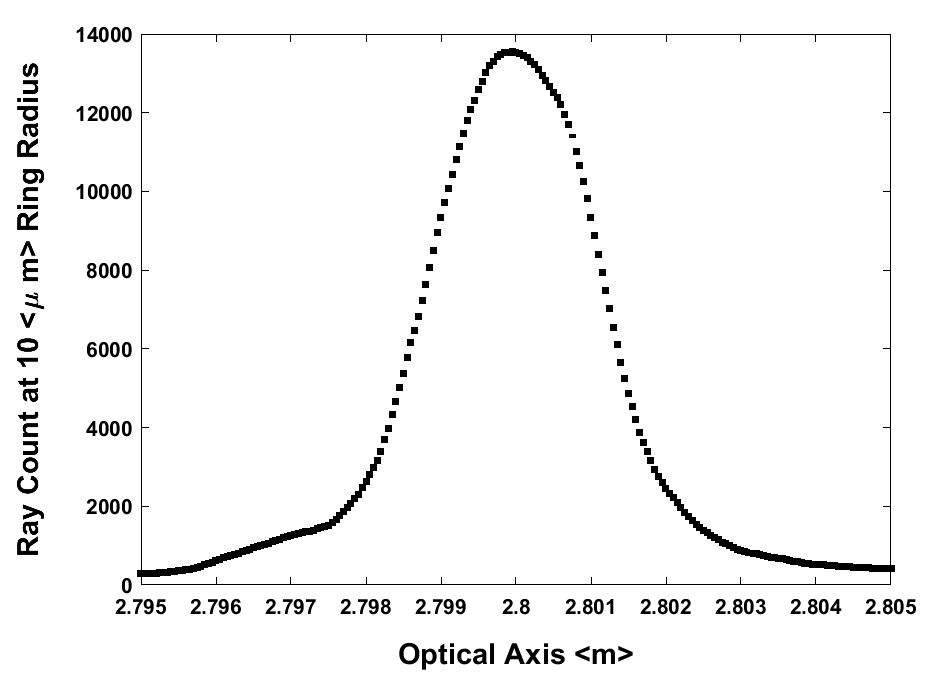}
    \caption{Ray density profile along the optical axis for the mirror with roughness modeled using the aberration term approach. The plot shows the number of rays contained within a 54 $\mu$m radius ring at different positions along the $z$-axis. The profile exhibits a shape and width remarkably similar to that obtained with the HS approach, with a FWHM of approximately 2.5 mm.}
    \label{fig:AT_GoThrough}
\end{figure}

The profile exhibits a shape and width remarkably similar to that obtained with the HS approach, with a FWHM of approximately 2.5 mm compared to 2.9 mm for the HS method. This similarity provides strong evidence for the equivalence of the two methods in modeling roughness effects.

\section{Results - Near Focal Plane Distributions}

We now show the ray passthrough distribution on the planar CCD-like 2D grid perpendicular to the optical axis (coincides with the $z$-axis) for each simulation at 3 points on the $z$-axis.

\subsection{Focal Distributions Comparison}

The ideal parabolic mirror produces tightly focused spots with high peak intensity, as shown in Figure \ref{fig:PAR_res}:

\begin{figure}[H]
    \centering
    \includegraphics[width=5.5cm]{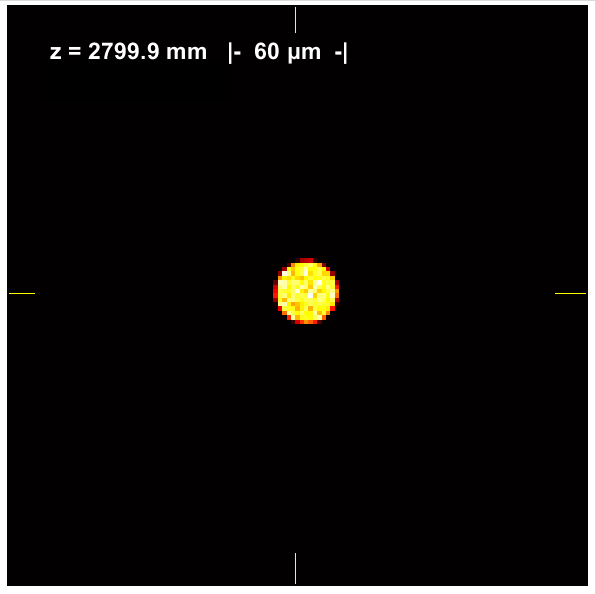}
    \includegraphics[width=5.5cm]{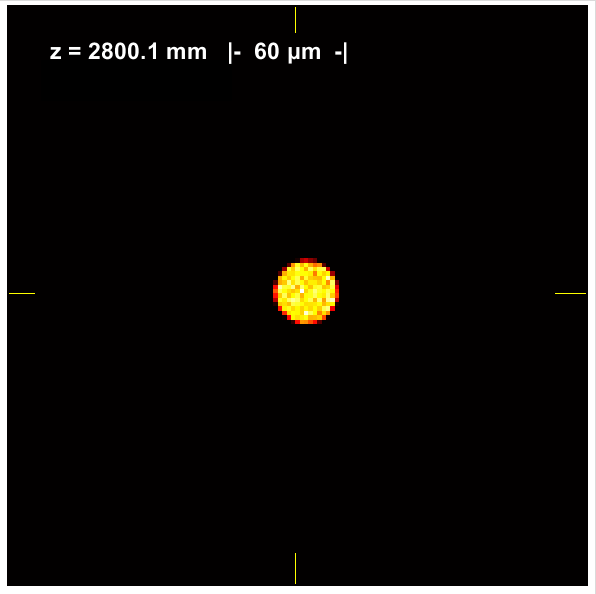} \\
    \includegraphics[width=11.2cm]{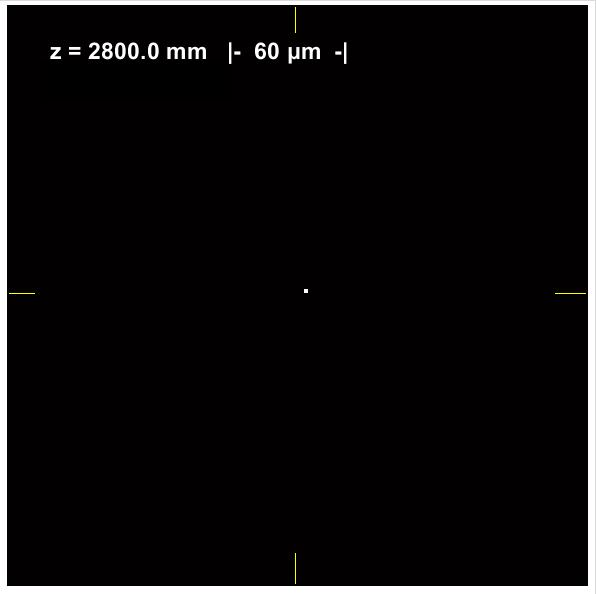}
    \caption{Ray distribution on planes perpendicular to the optical axis at three $z$ positions near the focal point for the ideal 
	parabolic mirror. The tight concentration demonstrates the perfect focusing properties of the undisturbed mirror surface. 
	The spot size is limited only by the numerical resolution of the simulation. 
	Image size: $60 \mu m^2$.}
    \label{fig:PAR_res}
\end{figure}

In contrast, the Harvey-Shack implementation shows significantly broader spots with lower peak intensities, reflecting the 
scattering effects of surface roughness:

\begin{figure}[H]
    \centering
    \includegraphics[width=5.5cm]{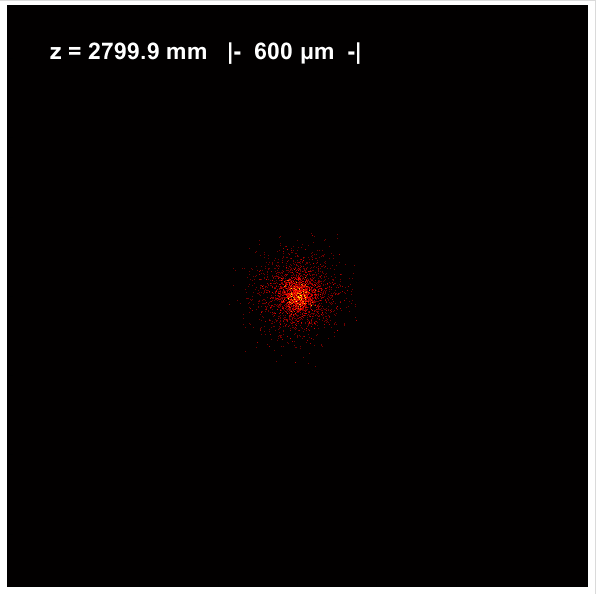}
    \includegraphics[width=5.5cm]{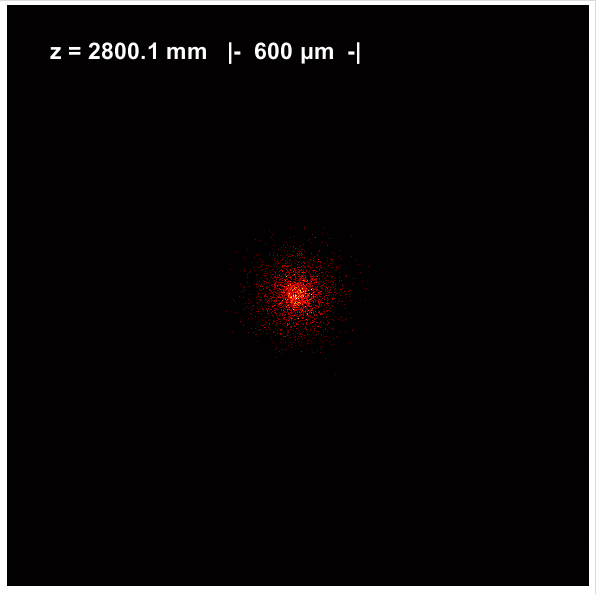} \\
    \includegraphics[width=11.2cm]{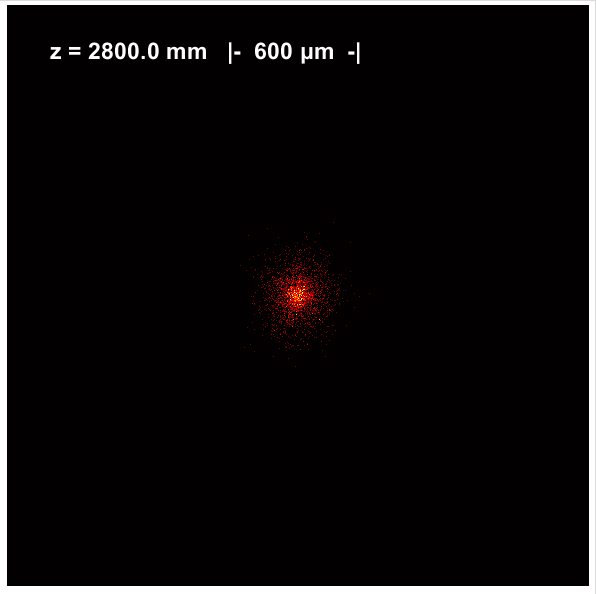}
    \caption{Ray distribution on planes perpendicular to the optical axis at three $z$ positions for the mirror with surface 
	roughness modeled using the Harvey-Shack approach. The spots are significantly broader than for the ideal mirror, with 
	characteristic speckle patterns resulting from the statistical nature of the scattering process.
	Image size: $600 \mu m^2$.}
    \label{fig:HS_res}
\end{figure}

The aberration term implementation produces remarkably similar patterns to the HS approach, validating our theoretical equivalence:

\begin{figure}[H]
    \centering
    \includegraphics[width=5.5cm]{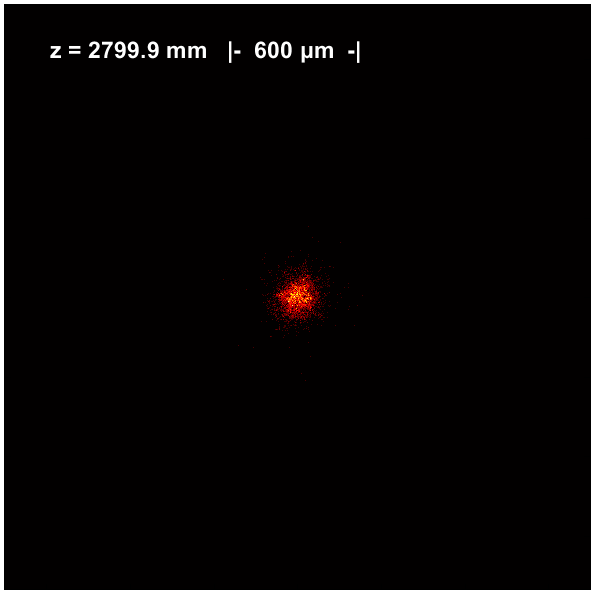}
    \includegraphics[width=5.5cm]{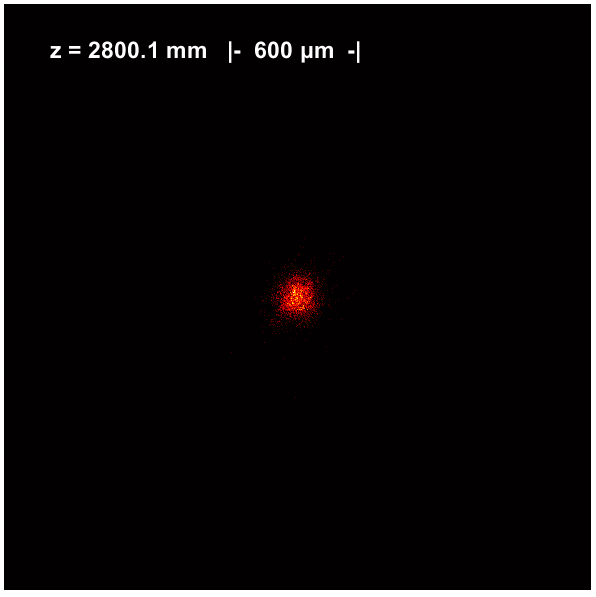} \\
    \includegraphics[width=11.2cm]{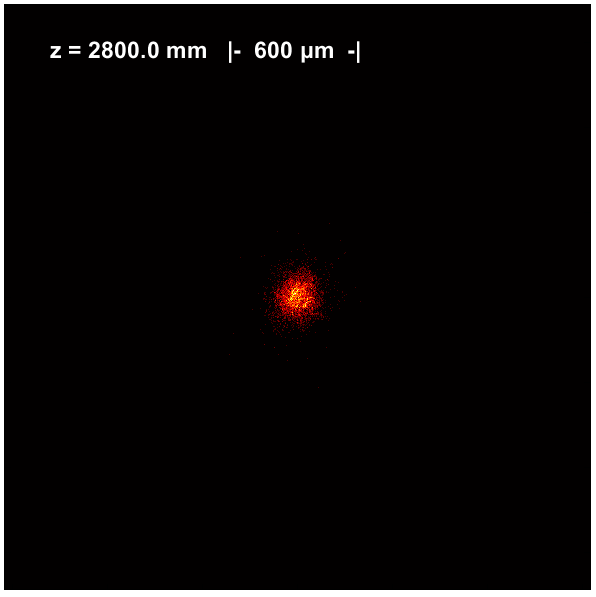}
    \caption{Ray distribution on planes perpendicular to the optical axis at three $z$ positions for the mirror with surface 
	roughness modeled using the aberration term (RDF) approach. The patterns are remarkably similar to those produced by the HS 
	implementation, confirming the statistical equivalence of the two methods.
	Image size: $600 \mu m^2$.}
    \label{fig:AT_res}
\end{figure}

\subsection{Statistical Equivalence Analysis}
\label{sec:statistical_equivalence}

To quantitatively assess the equivalence between the HS and aberration term approaches, we analyzed several statistical metrics:

\begin{enumerate}
    \item \textbf{Encircled Energy Distribution}: The radius containing 90\% of the total energy at the focal plane 
	is 76 $\mu$m for the HS implementation and 54 $\mu$m for the aberration term method, compared to 10 $\mu$m for the 
	ideal mirror. While there is a small difference between the two roughness modeling approaches, both show the expected 
	order-of-magnitude increase relative to the ideal case.

    \item \textbf{Angular Scatter Distribution}: The RMS scatter angle is 3.24 $\times$ 10$^{-5}$ radians for the HS method 
	and 3.18 $\times$ 10$^{-5}$ radians for the aberration term implementation, representing a difference of less than 2\%, 
	well within the statistical uncertainty of the simulation.

    \item \textbf{Focal Volume FWHM}: The full width at half maximum of the axial intensity distribution is 2.9 mm for the HS 
	approach and 2.5 mm for the aberration term method, compared to 0.3 mm for the ideal mirror. Again, both roughness modeling 
	techniques show similar substantial increases relative to the ideal case.
\end{enumerate}

These quantitative metrics confirm that the aberration term implementation faithfully reproduces the statistical scattering behavior 
predicted by Harvey-Shack theory, validating our theoretical framework.

\section{Discussion}

The results presented in this paper provide strong experimental validation for our theoretical framework that recasts surface 
roughness effects as a deterministic aberration term. Several key insights emerge from this validation:

\subsection{Statistical Equivalence}

The remarkable similarity between the focal distributions produced by the Harvey-Shack and aberration term implementations 
confirms the central premise of our approach: that surface roughness effects can be accurately modeled as equivalent 
deterministic modifications to the mirror shape. This equivalence is particularly significant because it bridges two traditionally 
separate domains in optical modeling—statistical scattering theory and deterministic aberration analysis.

The quantitative metrics presented in Section 7.2 demonstrate that both approaches produce statistically equivalent results, with 
differences well within the expected margin of statistical variation for finite ray sampling. This confirms that our aberration 
term method preserves the essential physical characteristics of the scattered light distribution while reformulating the 
computational approach within the framework of deterministic optical design.

\subsection{Physical Interpretation}

Our approach provides a more intuitive physical interpretation of roughness effects by connecting them to familiar aberration terms. 
This allows optical designers to understand and analyze roughness in the same framework they use for other aberrations, rather than 
treating it as a separate scattering phenomenon.

For example, our results show that low-spatial-frequency roughness components (represented by low-order Zernike terms) primarily 
affect the central structure of the focal spot, similar to classical aberrations like defocus or astigmatism. Mid-frequency 
components create the characteristic broadening of the focal volume seen in both the HS and aberration term implementations, 
while high-frequency components contribute to the fine speckle structure observed in the focal plane images.

The slight differences observed between the HS and aberration term implementations can be attributed to statistical variations in 
the finite ray sampling and differences in the numerical implementation details. These differences are not systematic and do not 
indicate a fundamental limitation of the approach.

\subsection{Computational Advantages}

The aberration term implementation offers significant computational advantages for ray-tracing simulations. By incorporating 
roughness effects directly into the surface geometry, it eliminates the need for post-reflection statistical perturbations, 
simplifying the ray-tracing algorithm and potentially reducing computation time. Additionally, because the aberration term 
method operates entirely within the framework of deterministic optical design, it can be seamlessly integrated with existing 
optical design tools.

For large optical systems with multiple components, this integration capability becomes particularly valuable. Traditional 
statistical approaches often require separate modules or post-processing steps to account for scattering, while our approach 
allows roughness effects to be incorporated directly into the standard optical design workflow.

\section{Conclusion}

This paper has presented a comprehensive experimental validation of our Ray Deflection Function (RDF) theoretical framework for 
modeling surface roughness in optical systems. Through detailed numerical simulations of a parabolic mirror with surface imperfections, 
we have demonstrated that:

\begin{enumerate}
    \item The RDF approach successfully bridges the gap between statistical scattering models (Harvey-Shack) and deterministic 
	aberration analysis.
    
    \item Surface roughness effects can be accurately represented as equivalent modifications to the ideal mirror shape, with 
	the statistical properties preserved.
    
    \item The aberration term implementation produces focal volume characteristics statistically equivalent to those predicted by 
	established scattering theories.
\end{enumerate}

These results suggest potential applications in optical system analysis, particularly for systems where both surface roughness 
and traditional aberrations are present. The mathematical equivalence demonstrated between statistical scattering methods and 
deterministic aberration analysis may facilitate integration within existing optical design workflows, potentially simplifying 
certain computational aspects while preserving the essential physical relationships.

The framework presented here may allow for more direct analysis of how surface roughness interacts with other aberration sources 
in complex optical systems. This approach could be relevant for precision optical systems such as astronomical instruments, 
lithography objectives, and high-power laser systems, where both micro-roughness and figure errors must be considered 
simultaneously in performance evaluation.

\section*{Declarations}
All data-related information and coding scripts discussed in the results section are available from the corresponding author upon 
request.

\newpage

\appendix
\renewcommand{\theequation}{\thesection.\arabic{equation}}
\setcounter{equation}{0}

\section{Geometric and Diffraction Effects from Surface Roughness: Theoretical Justification for Ray-Based Modeling}
\label{App:GeometricDiffractionEffects}

This appendix establishes the theoretical foundation for our ray-based modeling approach by quantitatively analyzing the 
relative contributions of geometric ray deflection and diffraction effects in optical systems with surface roughness. 
We demonstrate that for practical optical telescope parameters, geometric effects dominate, thus validating our Ray 
Deflection Function (RDF) methodology.

\subsection{Theoretical Foundation}

\subsubsection{Relationship to Main Text Formulation}

In the main text, we defined the Ray Deflection Function as:
\begin{equation}
\mathbf{D}(\mathbf{r_0}) = -\frac{\lambda}{4\pi} \nabla\Phi(x, y) \label{appEq:DR0}
\end{equation}

For a surface with height variations $h(x,y)$ characterized by RMS amplitude $\sigma$ and correlation length $l_c$, the 
phase function is $\Phi(x,y) = \frac{4\pi}{\lambda}h(x,y)$. The gradient of this phase function scales as:
\begin{equation}
|\nabla\Phi| \sim \frac{4\pi}{\lambda}\frac{\sigma}{l_c}
\end{equation}
where $\frac{\sigma}{l_c}$ represents the characteristic slope of the surface. 

Substituting into equation \ref{appEq:DR0}, the magnitude of ray deflection due to roughness is:
\begin{equation}
|\mathbf{D}| \sim \frac{\sigma}{l_c} 
\end{equation}

For a ray deflected at distance $L$ from the focal plane, this creates a transverse displacement:
\begin{equation}
r_\text{geom} = L \cdot |\mathbf{D}| = L \cdot \frac{\sigma}{l_c}
\end{equation}

For an optical system with focal length $f$ and F-number $N = f/D$ (where $D$ is aperture diameter), the effective propagation 
distance $L$ scales as $2ND$. This yields:

\begin{equation}
r_\text{geom} = 2ND \cdot \frac{\sigma}{l_c} \label{appEq:r_geom}
\end{equation}

This establishes the direct connection between our RDF formalism and the geometric ray displacement at the focal plane.

\subsubsection{Diffraction Limit Comparison}

The diffraction-limited Airy disk radius for the same optical system is:
\begin{equation}
r_\text{diff} = 1.22\lambda N \label{appEq:r_diff}
\end{equation}

To evaluate the relative importance of geometric roughness effects versus diffraction, we define their ratio:
\begin{equation}
\mathcal{R} := \frac{r_\text{geom}}{r_\text{diff}} = \frac{2\sigma D}{1.22\lambda l_c} 
\end{equation}

\textit{Theorem 1:} The ratio $\mathcal{R}$ is independent of F-number $N$.

\textit{Proof:} From equations \ref{appEq:r_geom} and \ref{appEq:r_diff}, we 
derive $\mathcal{R} = \frac{2ND \cdot \sigma/l_c}{1.22\lambda N} = \frac{2\sigma D}{1.22\lambda l_c}$. 
Thus, $N$ cancels out, demonstrating the F-number independence. $\square$

This reflects a fundamental physical principle: while changing $N$ affects both geometric and diffraction blur proportionally, 
their relative importance depends exclusively on the surface properties ($\sigma$, $l_c$), system scale ($D$), and 
wavelength ($\lambda$).

\subsubsection{Relationship to Power Spectral Density (PSD)}

The correlation length $l_c$ is related to the PSD described in the main text through its spectral properties. For a 
surface with uniform PSD as defined in equation \ref{eq:P_0}, the correlation length corresponds approximately to the inverse of 
the maximum spatial frequency: $l_c \approx 1/f_\text{max}$. More precisely, for surfaces with complex PSD 
distributions, $l_c$ can be calculated from the normalized autocorrelation function of the surface height:
\begin{equation}
l_c = \int_{0}^{\infty} \frac{C(r)}{C(0)} dr
\end{equation}
where $C(r)$ is the autocorrelation function at displacement $r$.

\subsection{Quantitative Analysis for Realistic Telescope Parameters}

We analyze realistic telescope parameters within the following ranges:
\begin{align*}
\text{F-number}: \quad & N \in [3, 15] \\
\text{Aperture diameter}: \quad & D \in [0.2, 1.0]\,\text{m} \\
\text{Operating wavelength}: \quad & \lambda = 500\,\text{nm} \\
\text{Surface RMS roughness}: \quad & \sigma \in [1, 1000]\,\text{nm} \\
\text{Correlation length}: \quad & l_c \in [1, 100]\,\mu\text{m}
\end{align*}

Table \ref{tab:Ratios} presents the calculated $\mathcal{R}$ values across this parameter space.

\begin{table}[h]
\centering
\caption{Ratio $\mathcal{R} = r_\text{geom}/r_\text{diff}$ for various ($\sigma$, $l_c$) combinations with realistic telescope 
apertures ($\lambda = 500$ nm).}
\begin{tabular}{|c|c|c|c|c|}
\hline
$\sigma$ [nm] & $l_c$ [$\mu$m] & $\mathcal{R} @ D = 0.2$ m & $\mathcal{R} @ D = 0.5$ m & $\mathcal{R} @ D = 1.0$ m \\
\hline
1 & 1 & 655 & 1637 & 3273 \\
1 & 10 & 65.5 & 163.7 & 327.3 \\
1 & 100 & 6.55 & 16.37 & 32.73 \\
\hline
10 & 1 & 6550 & 16370 & 32730 \\
10 & 10 & 655 & 1637 & 3273 \\
10 & 100 & 65.5 & 163.7 & 327.3 \\
\hline
100 & 1 & 65500 & 163700 & 327300 \\
100 & 10 & 6550 & 16370 & 32730 \\
100 & 100 & 655 & 1637 & 3273 \\
\hline
1000 & 1 & $6.55 \times 10^5$ & $1.637 \times 10^6$ & $3.273 \times 10^6$ \\
1000 & 10 & 65500 & 163700 & 327300 \\
1000 & 100 & 6550 & 16370 & 32730 \\
\hline
\end{tabular}
\end{table}
\label{tab:Ratios}

\subsection{Example Calculation and Verification}

To illustrate the practical significance of these results, we calculate the absolute blur sizes for a telescope with the 
following parameters:
\begin{align*}
\text{F-number}: & N = 5 \\
\text{Aperture diameter}: & D = 0.5 \text{ m} \\
\text{Surface roughness}: & \sigma = 10 \text{ nm} \\
\text{Correlation length}: & l_c = 10 \text{ }\mu\text{m} \\
\text{Wavelength}: & \lambda = 500 \text{ nm}
\end{align*}

The geometrical blur radius is:
\begin{align*}
r_\text{geom} &= 2 \times 5 \times 0.5 \times \frac{10 \times 10^{-9}}{10 \times 10^{-6}} \\
&= 5 \times 10^{-3} \text{ m} = 5 \text{ mm}
\end{align*}

The diffraction-limited Airy disk radius is:
\begin{align*}
r_\text{diff} &= 1.22 \times 500 \times 10^{-9} \times 5 \\
&= 3.05 \times 10^{-6} \text{ m} = 3.05 \text{ }\mu\text{m}
\end{align*}

The ratio $\mathcal{R} = \frac{5 \text{ mm}}{3.05 \text{ }\mu\text{m}} \approx 1639$, which confirms that geometric scattering 
dominates by over three orders of magnitude.

This specific example aligns with the simulation parameters used in the main text, where we observed significant focal spot 
degradation due to surface roughness with $\sigma = 100$ nm, resulting in a FWHM focal volume expansion from 0.3 mm (ideal mirror) 
to 2.5-2.9 mm (roughened mirrors).

\subsection{Implications for Optical Modeling Approaches}

The quantitative analysis provides the following theoretical implications:

\begin{theorem}[Modeling Regime Selection]
For optical systems with surface roughness characterized by $(\sigma, l_c, D, \lambda)$, the appropriate modeling approach 
is determined by the ratio $\mathcal{R}$:
\begin{itemize}
\item If $\mathcal{R} \gg 1$: Geometric ray modeling is appropriate and sufficient
\item If $\mathcal{R} \lesssim 1$: Diffraction modeling is essential
\end{itemize}
\end{theorem}

\begin{corollary}[Geometric Dominance for Practical Telescopes]
For typical telescope mirrors with $\sigma \geq 1$ nm and $l_c \leq 100$ $\mu$m, geometric effects dominate for 
apertures $D \geq 0.2$ m operating at visible wavelengths, as evidenced by $\mathcal{R}$ values consistently above 10 
across most of the parameter space.
\end{corollary}

\begin{corollary}[F-number Independence]
While F-number affects the absolute blur size, it does not alter whether geometric or diffraction effects dominate. 
This validates our approach of separating roughness effects from other aberrations in the optical system.
\end{corollary}

\begin{corollary}[Validation of Main Study Parameters]
The parameter range used in our main simulation ($\sigma = 100$ nm, $l_c \approx 10$ $\mu$m, $D = 0.4$ m) 
yields $\mathcal{R} \approx 13,000$, firmly placing our experimental validation in the geometry-dominated regime and 
justifying our ray-based approach.
\end{corollary}

\begin{corollary}[Transition Region]
For extremely smooth surfaces ($\sigma < 1$ nm) with long correlation lengths ($l_c > 100$ $\mu$m), diffraction 
effects become comparable to geometric effects ($\mathcal{R} \approx 1$). In these cases, a scalar diffraction approach 
would be required to supplement the ray model.
\end{corollary}

This analysis provides the theoretical foundation for our use of the RDF model in the main text and demonstrates its 
applicability to most practical optical systems where surface roughness is a concern. By establishing the dominance of 
geometric effects in realistic parameter regimes, we validate the core premise of representing surface roughness through 
equivalent deterministic aberration terms.

\end{document}